\documentclass{article} 
\usepackage[final]{colm2025_conference}

\usepackage{microtype}
\usepackage{hyperref}
\usepackage{url}
\usepackage{booktabs}

\usepackage{lineno}

\usepackage{xcolor} 
\usepackage{booktabs} 
\usepackage{graphicx} 
\usepackage{amsmath}
\usepackage{array}
\usepackage{multirow}
\usepackage{tcolorbox}
\usepackage{algorithm}
\usepackage{algorithmic}

\definecolor{darkblue}{rgb}{0, 0, 0.5}
\hypersetup{colorlinks=true, citecolor=darkblue, linkcolor=darkblue, urlcolor=darkblue}
\usepackage{tcolorbox}

\definecolor{darkblue}{rgb}{0, 0, 0.5}
\hypersetup{colorlinks=true, citecolor=darkblue, linkcolor=darkblue, urlcolor=darkblue}

\title{Can LLMs Handle WebShell Detection? Overcoming Detection Challenges with Behavioral Function-Aware Framework}

\author{%
  Feijiang Han\\
  University of Pennsylvania \\
  \texttt{feijhan@seas.upenn.edu}
  \And
  Jiaming Zhang \& Chuyi Deng \\
  Central South University \\
  \And
  Jianheng Tang ~ \& Yunhuai Liu \\
  Peking University \\
}

\begin{document}

\ifcolmsubmission
\linenumbers
\fi

\maketitle

\begin{abstract}
WebShell attacks---where adversaries implant malicious scripts on web servers---remain a persistent threat. Prior machine-learning and deep-learning detectors typically depend on task-specific supervision and can be brittle under data scarcity, rapid concept drift, and out-of-distribution (OOD) deployment. Large language models (LLMs) have recently shown strong code understanding capabilities, but their reliability for WebShell detection remains unclear.
We address this gap by (i) systematically evaluating seven LLMs (including GPT-4, LLaMA-3.1-70B, and Qwen-2.5 variants) against representative sequence- and graph-based baselines on 26.59K PHP scripts, and (ii) proposing \emph{Behavioral Function-Aware Detection} (BFAD), a behavior-centric framework that adapts LLM inference to WebShell-specific execution patterns.
BFAD anchors analysis on security-sensitive PHP functions via a \emph{Critical Function Filter}, constructs compact LLM inputs with \emph{Context-Aware Code Extraction}, and selects in-context demonstrations using \emph{Weighted Behavioral Function Profiling} (WBFP), which ranks examples by a behavior-weighted, function-level similarity. Empirically, we observe a consistent precision--recall asymmetry: larger LLMs often achieve high precision but miss attacks (lower recall), while smaller models exhibit the opposite tendency; moreover, off-the-shelf LLM prompting underperforms established detectors. BFAD substantially improves all evaluated LLMs, boosting F1 by 13.82\% on average; notably, GPT-4, LLaMA-3.1-70B, and Qwen-2.5-Coder-14B exceed prior SOTA benchmarks, while Qwen-2.5-Coder-3B becomes competitive with traditional methods. Overall, our results clarify when LLMs succeed or fail on WebShell detection, provide a practical recipe and highlight future directions for making LLM-based detection more reliable.
\end{abstract}

\section{Introduction}

The rapid growth of web applications and cloud services has expanded the attack surface of modern systems, making WebShells a persistent threat. A WebShell is a malicious script implanted on a web server that enables remote command execution, data exfiltration, and broader system compromise~\citep{starov2016no, cui2018WebShell,han2026beyonddetection}. A Cisco Talos report~\citep{talos2024} highlights the prevalence of such activity: web shells were observed in 35\% of incidents in Q4 2024, up from 10\% in the previous quarter. WebShells are particularly difficult to defend against because they are continually adapted---often via obfuscation or encryption---to evade conventional detection pipelines~\citep{hannousse2021handling}.

To counter this threat, prior work spans signature/heuristic defenses and learned detectors. Rule-based approaches are increasingly brittle given the diversity and rapid evolution of WebShell variants~\citep{le2021efficient, jinping2020mixed}. Learning-based methods, including deep models~\citep{pu2022bert}, can capture richer patterns but typically require substantial labeled data and costly retraining---resources that are often scarce or sensitive in practice~\citep{shang2024multi}. Moreover, these models can suffer from catastrophic forgetting under continual updates and may generalize poorly to novel obfuscation or encryption strategies~\citep{jinping2020mixed, zhang2025mmfdetect}.

Large language models (LLMs) have recently demonstrated strong performance on code-centric tasks, from generation to program understanding~\citep{ma2024large}, and have been explored for security applications such as vulnerability detection~\citep{liu2023WebShell, wang2025poster,han2026latex2layout}. With appropriate prompting, LLMs can be adapted to new tasks without additional training~\citep{nong2024chain, trad2025manual,han2026readbeforeyouthink}, and they can provide natural-language rationales that support analyst workflows~\citep{ma2024large}. However, their effectiveness for WebShell detection remains insufficiently characterized.

Applying LLMs to WebShell detection introduces challenges that are less pronounced in standard code analysis. WebShell payloads are often heavily obfuscated or encrypted and may be buried within predominantly benign code~\citep{liu2023WebShell}. Naively feeding an entire file to an LLM can therefore be ineffective: in our dataset, the longest WebShell contains 1,386,438 tokens, far beyond typical context windows, so truncation can easily remove the malicious core~\citep{wang2025poster, ceka2024can}. In addition, in-context learning (ICL) is fragile in this domain: the diversity of obfuscation patterns complicates demonstration selection, and demonstrations themselves consume context budget that would otherwise be allocated to the target code~\citep{yuan2024focused}. While longer-context LLMs are actively studied~\citep{chen2023extending}, prior work suggests that performance can degrade on longer inputs and latency may become prohibitive in practice~\citep{ma2024large, fang2024large}.

\textbf{In this paper, we take a step toward making LLM-based WebShell detection both measurable and reliable.}

\textbf{First,} we conduct a systematic evaluation of LLMs for WebShell detection and compare them with representative learned detectors. Concretely, we evaluate seven closed- and open-source LLMs spanning a wide range of scales---including GPT-4~\citep{achiam2023gpt}, LLaMA 3.1 70B~\citep{grattafiori2024llama}, Qwen 2.5 Coder (14B/3B)~\citep{yang2024qwen2}, and Qwen 2.5 (3B/1.5B/0.5B)~\citep{yang2024qwen2}---on a dataset of 26.59K PHP scripts (4.93K WebShells and 21.66K benign samples). For this comparison, we benchmark against several traditional methods, including GloVe+SVM~\citep{petridis2024text, rigutini2024performance}, CodeBERT+Random Forest~\citep{alghamdi2022comparative}, and graph-based approaches such as GAT~\citep{kang2023llm}.

Our experiments yield three main observations:
\begin{itemize}
    \item \textbf{Scale shifts the error mode:} larger LLMs (e.g., GPT-4, Qwen 2.5 Coder 14B) achieve very high precision but lower recall (e.g., GPT-4 recall 85.98\%), whereas smaller models often trade precision for recall (e.g., Qwen 2.5 Coder 3B precision 38.93\%).
    \item \textbf{Naive ICL is unreliable:} randomly selected demonstrations can degrade performance, and selecting demonstrations by semantic similarity alone yields limited improvements.
    \item \textbf{LLMs do not close the gap by default:} without task-specific adaptation, off-the-shelf LLM prompting remains behind strong learned detectors in overall F1. For example, the best-performing LLM baseline, Qwen 2.5 Coder 14B, achieves an F1 score of 96.39\%, but still trails the GAT-based detector (98.87\% F1).\footnote{While traditional models such as GAT can achieve higher accuracy, they typically require task-specific training and sustained data collection. In contrast, LLMs can be deployed via prompting, leveraging pretraining with minimal task-specific supervision.}
\end{itemize}

\textbf{Second,} we propose \emph{Behavioral Function-Aware Detection} (BFAD), a behavior-centric framework designed to make LLM inference effective under these constraints and to better balance precision and recall. BFAD couples risk-aware filtering with an enhanced ICL strategy that weights demonstrations by their alignment with discriminative malicious behaviors. Across all evaluated LLMs, BFAD improves F1 by 13.82\% on average; GPT-4 and Qwen 2.5 0.5B improve by 6.89\% and 51.23\%, respectively. For several models (including GPT-4, LLaMA 3.1 70B, Qwen 2.5 Coder 14B, and Qwen 2.5 Coder 3B), BFAD yields performance competitive with---and in some cases surpassing---strong traditional baselines.

\textbf{Finally,} we outline several forward-looking research directions suggested by our analysis, including (i) constructing privacy-preserving, large-scale synthetic benchmarks for stress-testing generalization, (ii) integrating graph-based behavioral representations with LLMs via multimodal alignment (e.g., graph encoders plus lightweight adapters), and (iii) designing agentic ``fast--slow'' detection pipelines with autonomous update loops to sustain performance under distribution shift.

To the best of our knowledge, this is the first work to systematically characterize when and why LLMs succeed or fail on WebShell detection, and to provide a practical framework that closes much of the gap to learned state-of-the-art detectors.
\section{Related Work}

\textbf{WebShell Detection Techniques.}
Early WebShell detection systems were predominantly rule-based, relying on signatures and handcrafted heuristics to flag malicious scripts~\citep{le2021efficient, jinping2020mixed}. While effective against known families, such approaches are brittle under obfuscation and rapid variant evolution because they depend on predefined patterns~\citep{hannousse2021handling}. Subsequent machine-learning (ML) methods introduced feature-based classification using textual or behavioral signals. For instance, \citet{jinping2020mixed} combined Random Forest and CNN models with N-gram and TF-IDF features, reporting strong performance on PHP WebShells but noting sensitivity to dataset balance and encrypted samples. Deep learning has further expanded the design space by leveraging semantic and structural representations, including transformer-based embeddings (e.g., CodeBERT)~\citep{pu2022bert} and graph neural networks for code structure modeling. \citet{zhang2025mmfdetect} proposed MMFDetect, which fuses CodeBERT-CL semantics with CNN-extracted visual features from RGB-mapped PHP code. Despite steady progress, supervised ML/DL detectors typically require substantial labeled data (often scarce in security settings) and can generalize poorly to heavily obfuscated or previously unseen WebShell variants, while incurring non-trivial training and maintenance costs~\citep{shang2024multi, jinping2020mixed}.

\textbf{LLMs for Code and Security Analysis.}
Large language models (LLMs) trained on massive code corpora have become a powerful interface for code understanding and generation, enabling applications ranging from synthesis~\citep{ma2024large} to vulnerability detection~\citep{sun2024llm4vuln} and reliability assessment~\citep{liu2024reliability}. In security contexts, LLMs are attractive because they support zero-shot and few-shot generalization via prompting and can provide natural-language rationales that improve analyst interpretability~\citep{nong2024chain,han2026zerotuning}. For example, \citet{ma2024large} showed that LLMs can generate evasive WebShells with carefully designed prompts, and \citet{sun2024llm4vuln} proposed retrieval-enhanced prompting for vulnerability reasoning. However, compared with these broader code-security applications, LLM-based WebShell detection remains relatively underexplored, especially for long, obfuscated scripts where malicious behavior may be sparsely embedded.

\textbf{Challenges in Applying LLMs to WebShell Detection.}
Directly applying LLMs to WebShell detection faces two practical bottlenecks. First, long WebShell files can exceed fixed context windows, causing truncation that may omit the truly malicious region~\citep{ceka2024can, wang2025poster}. Moreover, performance can degrade as input length grows; \citet{fang2024large} reported substantial accuracy drops for obfuscated long-context code. Techniques such as chunking and long-context model architectures (e.g., sparse attention) can partially alleviate this issue~\citep{guo2023longcoder, wang2024sparsecoder}, but may sacrifice global context and introduce fragmentation~\citep{wang2025poster}. Second, in-context learning (ICL) is itself context-expensive: demonstrations consume a significant fraction of the prompt budget~\citep{min2022rethinking, wang2025poster}, and naive retrieval (random or whole-file semantic similarity) often fails to capture the behavioral structure of WebShells. While information-theoretic criteria have been explored for ICL selection~\citep{liu2023towards}, existing formulations primarily target natural language classification and do not directly address behavior-centric code security tasks.

Taken together, the limitations of (i) rule-based rigidity, (ii) supervised ML/DL dependence on labeled data and weak OOD robustness, and (iii) LLM context and ICL constraints motivate BFAD. BFAD addresses long-context inputs via hybrid, behavior-anchored extraction and improves ICL reliability with WBFP, which retrieves demonstrations based on behavior-weighted, function-level profiles rather than generic similarity.

\section{Behavioral Function-Aware Detection Framework}

\begin{figure}[ht!]
    \centering \includegraphics[width=1\linewidth]{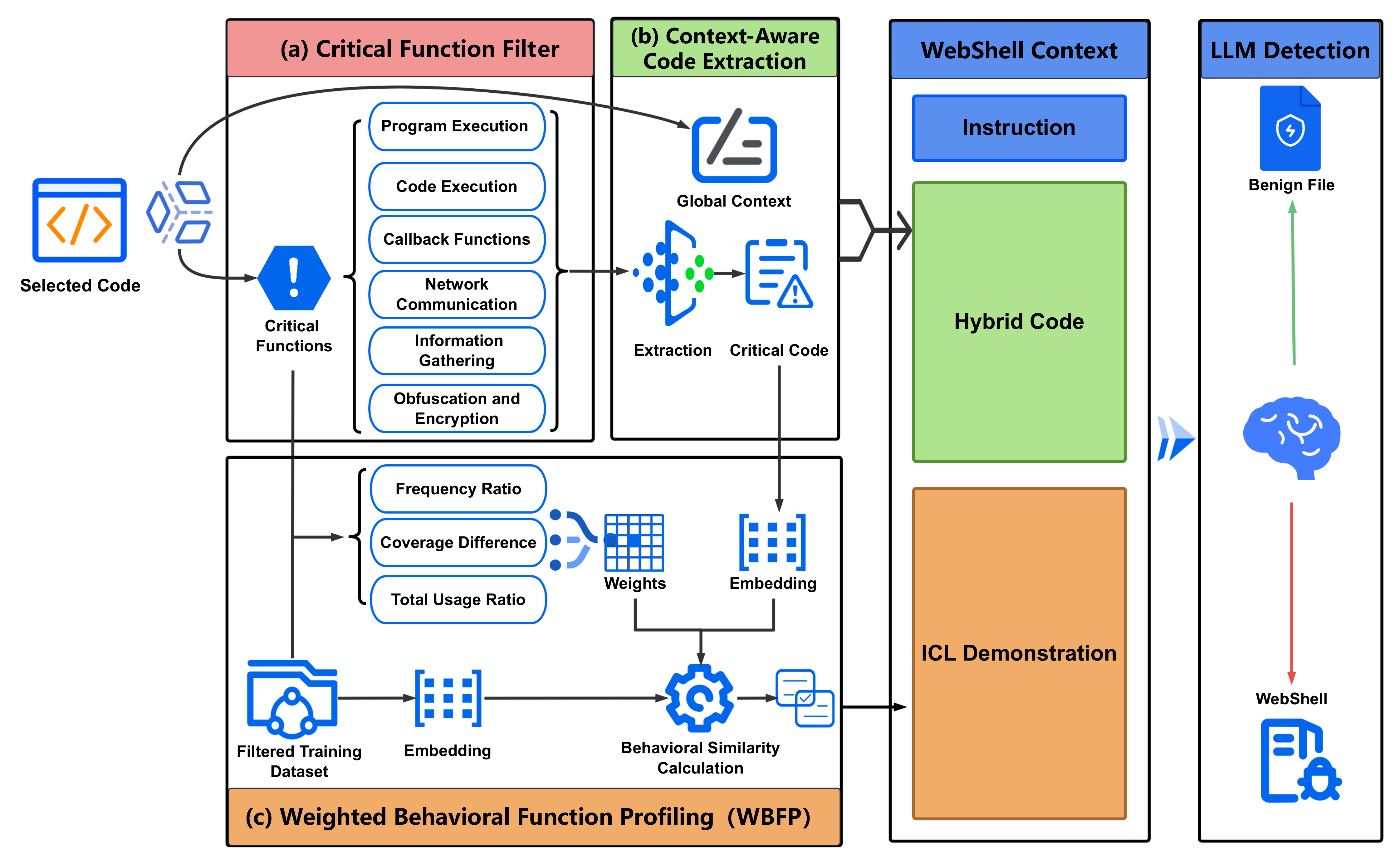}
    \caption{Overview of the Behavioral Function-Aware Detection framework for WebShell detection. It consists of three components: (a) Critical Function Filter, which identifies PHP functions associated with malicious behavior; (b) Context-Aware Code Extraction, which isolates critical code regions to overcome LLM context limitations; and (c) Weighted Behavioral Function Profiling, which selects ICL demonstrations using a behavior-weighted similarity score.}
    \label{fig:framework}
\end{figure}

We present \textbf{Behavioral Function-Aware Detection (BFAD)}, a framework that improves WebShell detection by (i) isolating behavior-relevant code snippets for LLM analysis and (ii) selecting in-context learning (ICL) demonstrations that better match the target file's malicious behaviors. As illustrated in Figure~\ref{fig:framework}, BFAD comprises three components: (a) a \textbf{Critical Function Filter} that maps PHP function calls to a behavior taxonomy and extracts behavior-relevant functions; (b) \textbf{Context-Aware Code Extraction} that constructs a compact yet informative code view under LLM context limits; and (c) \textbf{Weighted Behavioral Function Profiling} that ranks candidate demonstrations using a behavior-weighted similarity score computed from function-centric representations.

\subsection{Critical Function Filter}

WebShells often implement malicious behaviors---e.g., command execution, data exfiltration, and payload obfuscation---through a small set of high-risk PHP functions. In practice, these calls are frequently buried in long, noisy, and obfuscated code, which dilutes LLM attention and complicates robust detection. BFAD therefore begins with a \textbf{Critical Function Filter} that organizes PHP functions into six behavior categories: \textit{Program Execution}, \textit{Code Execution}, \textit{Callback Functions}, \textit{Network Communication}, \textit{Information Gathering}, and \textit{Obfuscation and Encryption}. This taxonomy captures the distinct roles that functions play in typical WebShell workflows.

Concretely, \textbf{Program Execution} includes functions such as \texttt{exec} and \texttt{system} that invoke system commands; \textbf{Code Execution} includes \texttt{eval} and \texttt{preg\_replace} (with the \texttt{/e} modifier in legacy PHP) that can interpret strings as code; and \textbf{Callback Functions} includes \texttt{array\_map} and \texttt{register\_shutdown\_function} that enable indirect invocation and are frequently used to hide control flow.

\textbf{Network Communication} includes \texttt{fsockopen} and \texttt{curl\_init} for remote communication (e.g., command-and-control or exfiltration); \textbf{Information Gathering} includes \texttt{phpinfo} and \texttt{getenv} for environment reconnaissance; and \textbf{Obfuscation and Encryption} includes \texttt{base64\_encode} and \texttt{openssl\_encrypt} for disguising or encrypting payloads.

Our statistical analysis (Appendix~\ref{app:critical_functions}) indicates that WebShell files invoke these critical functions far more frequently than benign files: on average, WebShells contain 22.76 critical-function calls, compared to 0.74 in benign files. This gap suggests strong behavioral signals, but it is insufficient by itself because benign programs may legitimately use the same APIs. We therefore treat critical functions as anchors to locate behavior-relevant code and rely on LLMs to disambiguate benign versus malicious intent from context.

\subsection{Context-Aware Code Extraction}

Building on the Critical Function Filter, BFAD performs \textbf{Context-Aware Code Extraction} to construct an LLM-friendly representation of each file. The key idea is to center extraction around critical function occurrences and include their surrounding context, so that the LLM receives the minimal code necessary to infer the intent of each call.

The complete extraction procedure is formalized in Algorithm~\ref{alg:context_extraction}, which takes as input the source code $\mathcal{C}$, the list of critical functions $\mathcal{F}$, and the context window size $\tau$, and produces a set of extracted critical code regions $\mathcal{R}$. 

We reduce input length by extracting windows around critical calls and merging overlapping windows, which preserves local behavioral evidence while avoiding redundant tokens. This operation can remove global context and over-emphasize critical functions, potentially increasing false positives for benign scripts that legitimately call high-risk APIs. To mitigate this, when the context budget allows we append additional truncated, non-overlapping segments from the remaining code, providing complementary global signals without exceeding the model's context limit.

\begin{algorithm}[h!]
\caption{Context-Aware Code Extraction}
\label{alg:context_extraction}
\begin{minipage}{\textwidth}
\begin{algorithmic}[1]
\STATE \textbf{Input:} Source code $\mathcal{C}$, list of critical functions $\mathcal{F}$, context window size $\tau$
\STATE \textbf{Output:} Extracted critical code regions $\mathcal{R}$
\STATE Initialize empty set of regions: $\mathcal{R} \gets \emptyset$
\FOR{each function $f \in \mathcal{F}$}
    \STATE Locate all occurrences of $f$ in $\mathcal{C}$
    \FOR{each occurrence of $f$ at position $p$}
        \STATE Extract context window $[p - \tau, p + \tau]$ from $\mathcal{C}$
        \STATE Add the extracted region to $\mathcal{R}$
    \ENDFOR
\ENDFOR
\STATE Merge overlapping regions in $\mathcal{R}$
\STATE Compute remaining context budget $B$
\IF{$B > 0$}
    \STATE Select additional non-overlapping code segments from $\mathcal{C}$
    \STATE Add selected segments to $\mathcal{R}$
\ENDIF
\STATE \textbf{return} $\mathcal{R}$
\end{algorithmic}
\end{minipage}
\end{algorithm}

\subsection{Weighted Behavioral Function Profiling}

Given extracted regions, BFAD selects ICL demonstrations via \textbf{Weighted Behavioral Function Profiling (WBFP)}. WBFP ranks candidate examples by a behavior-weighted similarity that emphasizes function types that best discriminate WebShells from benign files. Specifically, for each critical function type $f$, we compute three statistics from the WebShell and benign corpora: coverage difference ($r_c$), frequency ratio ($r_f$), and usage ratio ($r_u$). Coverage difference measures how often $f$ appears at least once in a file across the two corpora; frequency ratio measures the ratio of the average per-file call count; and usage ratio measures the ratio of the total call counts. We combine them into a discrimination score:

\[ \text{Score}_f = (r_c \cdot \alpha) + (r_f \cdot \beta) + (r_u \cdot \gamma), \]
where $\alpha$, $\beta$, and $\gamma$ are scalar hyperparameters (set to 1 in our experiments for equal contribution; see Section~\ref{sec:icl_settings}). We normalize scores into weights:

\[ w_f = \frac{\text{Score}_f}{\sum_{f' \in \mathcal{F}} \text{Score}_{f'}}.
\]

WBFP then computes similarity using embeddings $E(\cdot)$ from \textit{st-codesearch-distilroberta-base}~\citep{abi2023ml,al2023stacc}. Let $\mathcal{F}$ denote the set of critical function types. For each $f \in \mathcal{F}$, we concatenate the extracted regions associated with $f$ (denoted $R_f(x)$) and embed them as

\[ \mathbf{e}_f(x) = E(\text{concat}_f(x)). \]
The function-type similarity between files $x$ and $y$ is computed by cosine similarity:

\[ s_f(x, y) = \frac{\mathbf{e}_f(x) \cdot \mathbf{e}_f(y)}{\|\mathbf{e}_f(x)\| \|\mathbf{e}_f(y)\|}.
\]

The final similarity between the files is the weighted sum of the similarities:

\[ \text{Sim}(x, y) = \sum_{f \in \mathcal{F}} w_f \cdot s_f(x, y). \] 

This formulation prioritizes function types that are more indicative of WebShell behavior and down-weights less informative semantics, yielding demonstrations that better match the target file's behavioral profile.

\subsection{LLM-Based Detection Framework}

We integrate BFAD into an LLM-based detection pipeline that combines context-aware extraction with behavior-weighted demonstration selection. By providing behavior-relevant snippets together with complementary global context and a closely matched demonstration, the LLM is better positioned to distinguish malicious intent from benign uses of similar APIs.

The LLM input consists of (a) a \textbf{system directive} specifying the cybersecurity-analysis role and (b) a \textbf{user query} containing the extracted code view and the selected ICL demonstration. To balance effectiveness and cost, we include a single demonstration per query. The full prompt template is provided in Appendix~\ref{app:prompt}.
\section{Experiment}

\subsection{Dataset Overview}

We construct a dataset of 26,594 PHP scripts, including 21,665 benign programs and 4,929 WebShells. Benign programs are sourced from established open-source PHP projects to reflect realistic development practices. WebShells are collected from public security repositories and further augmented with synthetic obfuscation to increase behavioral and lexical diversity.\footnote{We acknowledge the risk of data leakage due to LLM pre-training. While pre-training does not explicitly optimize for WebShell classification, memorization of code snippets could still affect evaluation. To reduce overlap with potential training corpora, we restrict benign projects to those updated between October 2024 and 2025.}
To characterize the distributional shift and the challenge posed by long contexts, we measure sequence lengths using the GPT-4 tokenizer. WebShells are substantially longer than benign files, with a maximum of 1,386,438 tokens and an average of 30,856.60 tokens, compared to a maximum of 305,670 and an average of 2,242.89 tokens for benign programs. A detailed breakdown of the dataset is provided in Table~\ref{apptab:dataset_summary} (Appendix~\ref{app:dataset}).

\subsection{Experiment Setup}

\paragraph{ICL Settings.}
\label{sec:icl_settings}
We randomly sample 60\% of the dataset to form a fixed demonstration library for ICL. Using this subset, we compute WBFP-based weights for each critical function type by combining coverage difference ($r_c$), frequency ratio ($r_f$), and usage ratio ($r_u$). The resulting normalized weights quantify each function type's discriminative importance for separating WebShells from benign programs and are used to rank candidate demonstrations. The full weights are reported in Table~\ref{apptab:key_function_weights} (Appendix~\ref{app:dataset}).

\paragraph{Baseline Models, Hyperparameter Settings, and Evaluation Metrics.}
We compare BFAD-enhanced LLMs against representative classical and graph-based detectors: GloVe+SVM, CodeBERT+Random Forest, GCN, and GAT. For GloVe+SVM, we use 300-dimensional pre-trained GloVe embeddings and an SVM classifier with default hyperparameters~\citep{qi2018end, zeng2025WebShell}. For CodeBERT+Random Forest, we use CodeBERT embeddings (768 dimensions) with a Random Forest classifier using default settings~\citep{wang2024WebShell}. For the graph models, we follow prior work~\citep{feng2024glareshell}: both GCN and GAT are trained for 120 epochs with learning rate 0.001; GCN uses 3 hidden layers with hidden size 32, and GAT uses 3 hidden layers with hidden size 8 and 8 attention heads.
We report standard classification metrics: accuracy, precision, recall, and F1.
\section{Results and Analysis}

This section evaluates LLMs of different scales for WebShell detection and quantifies the gains delivered by our proposed BFAD framework. We structure the analysis around three research questions (RQs), covering both out-of-the-box LLM behavior and the contribution of individual BFAD components:

\begin{itemize}
\item \textbf{RQ1:} How do large and small-scale LLMs perform in WebShell detection compared to traditional ML and DL methods, and how does BFAD improve their effectiveness?
\item \textbf{RQ2:} How effective is context-aware code extraction at balancing global context and local behavioral focus under LLM context length constraints?
\item \textbf{RQ3:} How does WBFP improve demo selection for ICL?
\end{itemize}

\subsection{Performance Evaluation of LLMs and BFAD Enhancements (RQ1)}

To address RQ1, we compare seven LLMs with representative classical detectors. The LLMs include large-scale models (GPT-4, LLaMA-3.1-70B, Qwen-2.5-Coder-14B) and smaller models (Qwen-2.5-Coder-3B, Qwen-2.5-3B, Qwen-2.5-1.5B, Qwen-2.5-0.5B). Baselines include sequence-based methods (GloVe+SVM, CodeBERT+RF) and graph-based methods (GCN, GAT). Table~\ref{apptab:enhanced_prompt} reports all metrics.

\begin{table*}[h!]
\centering
\caption{Performance Comparison of BFAD-Enhanced Models Against Baselines.}
\label{apptab:enhanced_prompt}
\resizebox{\textwidth}{!}{
\begin{tabular}{@{} l l llll@{}}
\toprule
\textbf{Category} & \textbf{Model} & \textbf{Accuracy} & \textbf{Precision} & \textbf{Recall} & \textbf{F1 Score} \\
\midrule
\multirow{2}{*}{Sequence Baselines}
& GloVe+SVM & 96.20\% & 93.30\% & 94.30\% & 93.80\% \\
& CodeBERT+RF & 96.30\% & 94.00\% & 95.60\% & 94.80\% \\
\midrule
\multirow{2}{*}{Graph Baselines}
& GCN & 96.90\% & 94.40\% & 95.30\% & 94.90\% \\
& GAT & 98.37\% & 99.52\% & 97.39\% & 98.87\% \\
\midrule
\multirow{3}{*}{LLM Baselines (Large)}
& GPT-4 & 97.27\% & 100.00\% & 85.98\% & 92.46\% \\
& LLaMA-3.1-70B & 98.01\% & 97.31\% & 92.36\% & 94.77\% \\
& Qwen-2.5-Coder-14B & 98.64\% & 99.32\% & 93.63\% & 96.39\% \\
\midrule
\multirow{4}{*}{LLM Baselines (Small)}
& Qwen-2.5-Coder-3B & 71.11\% & 38.93\% & 99.32\% & 55.93\% \\
& Qwen-2.5-3B & 93.72\% & 78.03\% & 91.84\% & 84.37\% \\
& Qwen-2.5-1.5B & 43.62\% & 34.61\% & 95.77\% & 50.84\% \\
& Qwen-2.5-0.5B & 19.47\% & 18.65\% & 100.00\% & 31.44\% \\
\midrule
\multirow{7}{*}{LLM + BFAD}
&  GPT-4 & 99.75\% & 100.00\% & 98.71\% & 99.35\% \textcolor[rgb]{0,0.5,0}{(+6.89)} \\
&  LLaMA-3.1-70B & 99.38\% & 98.72\% & 98.09\% & 98.40\% \textcolor[rgb]{0,0.5,0}{(+3.63)} \\
&  Qwen-2.5-Coder-14B & 98.76\% & 98.68\% & 94.90\% & 96.75\% \textcolor[rgb]{0,0.5,0}{(+0.36)} \\
& Qwen-2.5-Coder-3B & 78.89\% & 46.67\% & 100.00\% & 63.64\% \textcolor[rgb]{0,0.5,0}{(+7.71)} \\
& Qwen-2.5-3B & 97.39\% & 88.64\% & 99.36\% & 93.69\% \textcolor[rgb]{0,0.5,0}{(+9.32)} \\
& Qwen-2.5-1.5B & 80.40\% & 48.51\% & 100.00\% & 65.33\% \textcolor[rgb]{0,0.5,0}{(+14.49)} \\
& Qwen-2.5-0.5B & 91.94\% & 71.10\% & 98.73\% & 82.67\% \textcolor[rgb]{0,0.5,0}{(+51.23)} \\
\bottomrule
\end{tabular}
}
\end{table*}

\paragraph{Performance Summary}
Our results highlight two key observations. First, vanilla LLM prompting exhibits a pronounced, scale-dependent precision--recall trade-off and remains below the strongest graph baseline (GAT). For example, GPT-4 attains 100\% precision but only 85.98\% recall, whereas Qwen-2.5-0.5B achieves 100\% recall with only 18.65\% precision. This asymmetry suggests that off-the-shelf LLMs are poorly calibrated for WebShell detection, motivating behavior-aware intervention.

Second, BFAD improves every evaluated LLM, elevating large models beyond the GAT baseline and making small models substantially more competitive. With BFAD, GPT-4 improves by +6.89 F1 to 99.35\%, surpassing GAT. The largest relative gain occurs for Qwen-2.5-0.5B (+51.23 F1), indicating that explicitly steering attention toward behavior-relevant code and demonstrations can unlock strong performance without task-specific fine-tuning.

\paragraph{Analysis of Model Behavior: Misclassification Patterns}


The above gap is consistent with qualitative misclassification patterns. Smaller models (e.g., Qwen-2.5-0.5B) often rely on surface heuristics: they over-trigger on high-risk API names such as \texttt{base64\_decode} and ignore surrounding intent, yielding high recall but many false positives when such APIs appear in benign utilities (e.g., cryptography libraries). In contrast, larger models (e.g., LLaMA-3.1-70B) more frequently leverage broader context and therefore maintain high precision, but can be distracted by plausible-looking surrounding logic and miss carefully disguised WebShells, producing false negatives. BFAD mitigates both failure modes by anchoring analysis on behavior-critical calls while still preserving enough local (and partial global) context for intent disambiguation.

\paragraph{Analysis of Model Behavior: Counterintuitive Scaling}

We observe a counterintuitive scaling pattern: Qwen-2.5-1.5B, while stronger than the 0.5B variant under vanilla prompting, improves less under BFAD and remains worse than Qwen-2.5-0.5B with BFAD. A plausible explanation is scale-dependent sensitivity to ICL. Very small models are more prone to label bias and may over-imitate the demonstration label rather than perform independent reasoning~\citep{fei2023mitigating}; when WBFP retrieves well-matched demonstrations, this bias can translate into large gains. As model size increases, the model may rely more on pre-training priors and partially resist demonstration guidance, which can introduce conflicts and reduce the net benefit~\citep{peng2025enhancing}. Thus, the strong BFAD performance of the 0.5B model should be interpreted as a demonstration-following effect rather than superior underlying reasoning.


\subsection{The Effectiveness of Context-Aware Code Extraction (RQ2)}

To answer RQ2, we evaluate Context-Aware Code Extraction on a large model (GPT-4) and a smaller model (Qwen-2.5-3B). We compare three input configurations: (1) the full source code, (2) extracted critical regions only, and (3) a hybrid input that concatenates critical regions with additional truncated code segments. Results are reported in Tables~\ref{apptab:cace_results_small} and~\ref{apptab:cace_results_large} (Appendix~\ref{app:results}).

\paragraph{Impact of Critical Regions}
For Qwen-2.5-3B, focusing on critical regions substantially improves performance over full-code prompting. At $\tau=100$, F1 increases from 84.37\% to 90.91\% (+6.54), with precision increasing from 78.03\% to 86.71\% (+8.68) and recall increasing from 91.84\% to 95.54\% (+3.70). This supports the hypothesis that removing irrelevant context sharpens behavioral evidence for smaller models. However, increasing $\tau$ can re-introduce noise: at $\tau=300$, F1 drops to 88.17\%, consistent with limited robustness to longer, noisier contexts.

For GPT-4, critical regions mainly improve recall with a minor precision trade-off. At $\tau=300$, recall rises from 85.98\% to 96.18\% (+10.20), while precision decreases from 100.00\% to 98.69\% (-1.31), resulting in an F1 increase from 92.46\% to 97.42\% (+4.96). This indicates that localized behavioral cues are particularly helpful for recovering missed attacks, while removing global context can slightly increase false positives.

\paragraph{Balancing Precision and Recall with the Hybrid Strategy}

The hybrid strategy improves the precision--recall balance by re-introducing limited global context. For Qwen-2.5-3B, it increases precision relative to critical-regions-only (86.71\% $\rightarrow$ 89.02\% at $\tau=100$; 82.32\% $\rightarrow$ 85.55\% at $\tau=300$). Although recall decreases slightly, the hybrid input yields a higher F1 at $\tau=300$ (89.70\%, +1.53 over 88.17), suggesting that carefully chosen global snippets reduce false positives when longer windows add noise. For GPT-4, the hybrid input improves recall without sacrificing precision: at $\tau=300$, recall increases from 85.98\% (full code) to 96.82\% (+10.84) while precision remains 100.00\%, yielding an F1 of 98.38\% (+5.92 over full code and +0.96 over critical-regions-only).

\subsection{The Effectiveness of WBFP for In-Context Learning (RQ3)}

To answer RQ3, we evaluate WBFP for ICL demonstration selection on Qwen-2.5-3B and GPT-4, using the best-performing hybrid input from RQ2. We compare four retrieval strategies: Random selection, source-code semantic similarity (SC-Sim), WBFP with equal weights (WBFP-Eq), and WBFP with function-level weights (WBFP-Wt). Results are reported in Tables~\ref{apptab:icl_results_small} and~\ref{apptab:icl_results_large} (Appendix~\ref{app:results}).

\paragraph{Limitations of Random and Semantic Similarity-Based Selection}

Random demonstrations substantially degrade ICL performance, indicating that irrelevant examples introduce strong prompt noise. For Qwen-2.5-3B ($\tau=100$), F1 drops to 60.83\% (from 90.97\% without ICL), driven primarily by a precision collapse (89.02\% $\rightarrow$ 46.33\%) while recall remains high (88.53\%). For GPT-4 ($\tau=300$), F1 decreases to 76.22\% (from 98.38\%), largely due to reduced precision (100.00\% $\rightarrow$ 65.58\%). These results show that naive ICL can be harmful without behaviorally relevant retrieval.

SC-Sim, which computes semantic similarity over entire source files, is also suboptimal. For Qwen-2.5-3B, SC-Sim yields 84.36\% F1 (vs.\ 90.97\% without ICL), with precision 78.57\% and recall 91.08\%. For GPT-4, SC-Sim reaches 96.32\% F1 with perfect precision but reduced recall (92.90\% vs.\ 96.82\%). We hypothesize this occurs because whole-file similarity is dominated by behavior-irrelevant regions, which dilutes the behavioral cues needed for effective WebShell retrieval.

\paragraph{Superiority of WBFP in Demonstration Selection}

WBFP improves ICL reliably, and WBFP-Wt is consistently the strongest variant. For Qwen-2.5-3B, WBFP-Wt achieves 93.69\% F1, improving over SC-Sim by +9.33 and over WBFP-Eq by +1.28. The gain comes from both higher precision (88.64\% vs.\ 78.57\% for SC-Sim) and near-perfect recall (99.36\%). These results suggest that behavior-weighted, function-centric retrieval is particularly beneficial for smaller models whose general code reasoning is limited.

For GPT-4, WBFP-Wt achieves 99.35\% F1 (+3.03 over SC-Sim and +0.34 over WBFP-Eq). Precision remains 100.00\% across WBFP variants, while WBFP-Wt improves recall to 98.71\% (vs.\ 92.90\% for SC-Sim and 98.03\% for WBFP-Eq). This indicates that behavior-aware retrieval primarily helps large models recover hard positives without sacrificing precision.

\section{Limitations and Future Work}
\label{sec:limitations}

In this section, we outline several concrete directions for future work---including synthetic benchmark construction, graph--LLM integration, and agentic detection pipelines---with the goal of making our ideas actionable and easy to build upon.

\subsection{Dataset Generalization}
A central challenge in WebShell research is the limited availability of diverse, real-world data. As in prior work, we primarily rely on public datasets, which may not fully reflect emerging, targeted, or highly obfuscated threats, and therefore limits our ability to rigorously assess generalization. An important direction is to use LLMs to \textbf{construct large-scale synthetic benchmarks} that are both privacy-preserving and diverse, enabling stress tests across a broader spectrum of attack vectors.

\subsection{Model and System Robustness}
Our framework is primarily static, and thus remains susceptible to advanced obfuscation and previously unseen (e.g., zero-day) behaviors. Future work should improve both robustness and adaptability. To mitigate evasive variants, one could incorporate \textbf{dynamic analysis}, e.g., by extracting runtime traces from sandboxed execution~\citep{zhao2024malicious,han2026beyonddetection}. 

To strengthen semantic understanding, \textbf{domain-specific fine-tuning} of code-oriented LLMs on curated WebShell corpora is another promising direction. Beyond standard fine-tuning, we propose a concrete next step that frames WebShell detection as a multimodal fusion problem: prior work suggests that graph representations can capture program behavior (e.g., function-call structure) more faithfully than pure sequence models~\citep{han2026beyonddetection}. Building on this observation, future work could (i) train a dedicated \textbf{graph encoder} to map code into behavior-centric embeddings (instead of relying solely on the LLM's text embeddings), and then (ii) learn an \textbf{adapter} that aligns the graph embeddings with natural-language instructions and the LLM's latent space (in the spirit of adapter-based alignment in VLMs and emerging graph-LLM systems). We believe this design can combine the strengths of both worlds: robust behavioral feature extraction from graphs, and the LLM's pretrained discrimination and interpretability.

For deployment, a practical direction is a multi-model \textbf{agentic} system that orchestrates specialized components. These components naturally fit a \textbf{multi-stage ``fast--slow'' pipeline}: a lightweight model performs high-recall triage, followed by a larger model that conducts deeper analysis on flagged samples. Finally, to sustain performance under distribution shift, the pipeline could be augmented with an \textbf{autonomous update loop} that discovers new patterns and refreshes detection models over time.

\section{Conclusion}

This paper provides a systematic study of LLMs for WebShell detection and shows that off-the-shelf prompting is not sufficient: across model scales, vanilla LLMs exhibit a pronounced precision--recall imbalance and fall short of strong supervised baselines. To bridge this gap, we introduce \textbf{Behavioral Function-Aware Detection (BFAD)}, which adapts LLM inference to WebShell behaviors through (i) critical-function anchoring, (ii) context-aware extraction that fits long scripts into a limited prompt budget, and (iii) behavior-weighted demonstration retrieval via WBFP. Empirically, BFAD consistently improves all evaluated LLMs and enables several models to match or surpass state-of-the-art detectors. Overall, our findings clarify both the failure modes and the promise of LLM-based WebShell detection, and suggest that behavior-centric prompting and retrieval are key to making LLMs reliable in security-critical code analysis.

\bibliography{colm2025_conference}
\bibliographystyle{colm2025_conference}

\newpage
\appendix
\onecolumn

\section{Prompt Details}
\label{app:prompt}

\begin{tcolorbox}[title={Prompt for WebShell Detection}, colframe=blue!50!black, colback=blue!10!white, top=1mm, bottom=1mm]
\textbf{System Prompt:} You are tasked with analyzing PHP scripts. Your objective is to classify the provided PHP code as either a WebShell or a legitimate script. A WebShell is typically a malicious script intended to exploit the server by executing unauthorized commands or providing backdoor access.

\textbf{User Prompt:} Analyze the provided PHP code to determine whether it constitutes a WebShell or a legitimate script. Provide your verdict as WebShell or benign.

\textbf{[Critical Code]} 

\textbf{[Source Code]} 

\textbf{[Examples]} 

\textbf{Output:}
\end{tcolorbox}

\newpage
\section{Critical Function Details}
\label{app:critical_functions}

\begin{table}[h!]
    \centering
    \caption{Statistics of Critical Functions in WebShell and Benign Programs. This table reports the percentage of files containing each function category and the average number of occurrences per file, with the ``Total'' row aggregating statistics across all categories.}
    \label{apptab:function_statistics}
    \renewcommand{\arraystretch}{1.2} 
    \begin{tabular}{@{} p{4cm} c c c @{}}
        \toprule
        \textbf{Function Category} & \textbf{Metric} & \textbf{WebShell Files} & \textbf{Normal Files} \\
        \midrule
        \multirow{2}{*}{Program Execution}
            & Files with Function (\%) & 53.06 & 1.54 \\
            & Avg. Occurrences per File & 3.21 & 0.03 \\
        \multirow{2}{*}{Code Execution}
            & Files with Function (\%) & 85.03 & 14.79 \\
            & Avg. Occurrences per File & 8.30 & 0.36 \\
        \multirow{2}{*}{Callback Functions}
            & Files with Function (\%) & 34.69 & 6.47 \\
            & Avg. Occurrences per File & 0.92 & 0.11 \\
        \multirow{2}{*}{Network Communication}
            & Files with Function (\%) & 50.34 & 2.77 \\
            & Avg. Occurrences per File & 1.69 & 0.04 \\
        \multirow{2}{*}{Information Gathering}
            & Files with Function (\%) & 46.26 & 2.77 \\
            & Avg. Occurrences per File & 5.46 & 0.05 \\
        \multirow{2}{*}{Obfuscation and Encryption}
            & Files with Function (\%) & 69.39 & 9.86 \\
            & Avg. Occurrences per File & 3.19 & 0.16 \\
        \midrule
        \multirow{2}{*}{\textbf{Total (All Functions)}}
            & Files with Function (\%) & 91.16 & 20.49 \\
            & Avg. Occurrences per File & 22.76 & 0.74 \\
        \bottomrule
    \end{tabular}
\end{table}

\newpage

\section{Dataset Details}
\label{app:dataset}

\begin{table*}[h!]
    \centering
    \caption{Dataset Composition, Distribution, and Sources. The dataset comprises 26,594 PHP scripts, categorized into benign programs and WebShells, with their respective counts, proportions, and sources.}
    \label{apptab:dataset_summary}
    \begin{tabular}{@{} >{\centering\arraybackslash}p{2cm} c >{\centering\arraybackslash}p{2cm} p{6.5cm} @{}}
        \toprule
        \textbf{Category} & \textbf{Count} & \textbf{Percentage} & \textbf{Source References} \\
        \midrule
        Benign Programs
            & 21,665 & 81.5\% & \href{https://github.com/getgrav/grav}{Grav}, \href{https://github.com/octobercms/october}{OctoberCMS}, \href{https://github.com/laravel/laravel}{Laravel}, \href{https://github.com/WordPress/WordPress}{WordPress}, \href{https://github.com/joomla/joomla-cms}{Joomla}, \href{https://github.com/nextcloud/server}{Nextcloud}, \href{https://github.com/symfony/symfony}{Symfony}, \href{https://github.com/bcit-ci/CodeIgniter}{CodeIgniter}, \href{https://github.com/yiisoft/yii2}{Yii2}, \href{https://github.com/cakephp/cakephp}{CakePHP}, \href{https://github.com/Intervention/image}{Intervention/Image}, \href{https://github.com/typecho/typecho}{Typecho} \\
        \midrule
        WebShells
            & 4,929 & 18.5\% & \href{https://github.com/xl7dev/WebShell}{WebShell}, \href{https://github.com/tanjiti/webshellSample}{WebshellSample}, \href{https://github.com/webshellpub/awsome-webshell}{Awsome-Webshell}, \href{https://github.com/DeEpinGh0st/PHP-bypass-collection}{PHP-Bypass-Collection}, \href{https://github.com/tdifg/WebShell}{WebShell (tdifg)}, \href{https://github.com/lhlsec/webshell}{Webshell (lhlsec)}, \href{https://github.com/bartblaze/PHP-backdoors}{PHP-Backdoors}, \href{https://github.com/tennc/webshell}{Tennc/Webshell}, \href{https://github.com/JohnTroony/php-webshells}{PHP-Webshells}, \href{https://github.com/BlackArch/webshells}{BlackArch/Webshells}, \href{https://github.com/JiaHeng-DLUT/Webshell-samples}{Webshell-Samples}, \href{https://github.com/leett1/Programe}{Programe}, \href{https://github.com/zhangchi991022/webshellDetection}{WebshellDetection}, \href{https://github.com/GabgM/WebShellCollection}{WebShellCollection}, \href{https://github.com/1337r0j4n/php-backdoors}{PHP-Backdoors (1337r0j4n)}, \href{https://github.com/Cyc1e183/PHP-Webshell-Dataset}{PHP-Webshell-Dataset}, \href{https://github.com/xiaoxiaoleo/xiao-webshell.git}{Xiao-Webshell} \\
        \midrule
        Total
            & 26,594 & 100.0\% & --- \\
        \bottomrule
    \end{tabular}
\end{table*}

\begin{table}[h!]
    \centering
    \caption{Normalized Scores for Key Function Categories. These scores reflect the weighted behavioral significance of each category as computed by the WBFP method.}
    \label{apptab:key_function_weights}
    \begin{tabular}{l c}
        \toprule
        \textbf{Function Category} & \textbf{Normalized Score} \\
        \midrule
        Program Execution          & 0.2068 \\
        Code Execution             & 0.2081 \\
        Callback Functions         & 0.0790 \\
        Network Communication      & 0.1498 \\
        Information Gathering      & 0.1861 \\
        Obfuscation and Encryption & 0.1702 \\
        \bottomrule
    \end{tabular}
\end{table}

\newpage

\section{Results}
\label{app:results}

\begin{table*}[h!]
    \centering
    \caption{Performance of Context-Aware Code Extraction with Qwen-2.5-3B with Different Context Lengths and Strategies.}
    \label{apptab:cace_results_small}
    \begin{tabular}{@{} l c c c c @{}}
        \toprule
        \textbf{Method} & \textbf{Accuracy} & \textbf{Precision} & \textbf{Recall} & \textbf{F1 Score} \\
        \midrule
        Source Code (Vanilla)         & 93.72\% & 78.03\% & 91.84\% & 84.37\% \\
        Critical Regions ($\tau=100$) & 96.28\% & 86.71\% & 95.54\% & 90.91\% \\
        Critical Regions ($\tau=200$) & 95.78\% & 84.75\% & 95.54\% & 89.82\% \\
        Critical Regions ($\tau=300$) & 95.04\% & 82.32\% & 94.90\% & 88.17\% \\
         \textbf{Hybrid Strategy ($\tau=100$)}  & \textbf{96.40}\% & \textbf{89.02}\% & \textbf{92.99}\% & \textbf{90.97}\% \\
        Hybrid Strategy ($\tau=200$)  & 95.66\% & 85.47\% & 93.63\% & 89.36\% \\
        Hybrid Strategy ($\tau=300$)  & 95.78\% & 85.55\% & 94.27\% & 89.70\% \\
        \bottomrule
    \end{tabular}
\end{table*}

\begin{table*}[h!]
    \centering 
    \caption{Performance of Context-Aware Code Extraction with GPT-4 with Different Context Lengths and Strategies.}
    \label{apptab:cace_results_large}
    \begin{tabular}{@{} l c c c c @{}}
        \toprule
        \textbf{Method} & \textbf{Accuracy} & \textbf{Precision} & \textbf{Recall} & \textbf{F1 Score} \\
        \midrule
        Source Code (Vanilla)         & 97.27\% & 100.00\% & 85.98\% & 92.46\% \\
        Critical Regions ($\tau=100$) & 98.51\% & 99.32\% & 92.99\% & 96.05\% \\
        Critical Regions ($\tau=200$) & 99.01\% & 99.34\% & 95.54\% & 97.40\% \\
        Critical Regions ($\tau=300$) & 99.01\% & 98.69\% & 96.18\% & 97.42\% \\
        Hybrid Strategy ($\tau=100$)  & 99.01\% & 100.00\% & 94.90\% & 97.39\% \\
        Hybrid Strategy ($\tau=200$)  & 99.14\% & 100.00\% & 95.81\% & 97.86\% \\
         \textbf{Hybrid Strategy ($\tau=300$)}  & \textbf{99.38}\% & \textbf{100.00}\% & \textbf{96.82}\% & \textbf{98.38}\% \\
        \bottomrule
    \end{tabular}
\end{table*}


\begin{table}[h!]
    \centering
    \caption{Comparison of Demonstration Selection Strategies for In-Context Learning with Qwen-2.5-3B (Under Best Hybrid Strategy $\tau=100$).}
    \label{apptab:icl_results_small}
    \begin{tabular}{@{} l c c c c @{}}
        \toprule
        \textbf{Method} & \textbf{Accuracy} & \textbf{Precision} & \textbf{Recall} & \textbf{F1 Score} \\
        \midrule
        No-ICL          & 96.40\% & 89.02\% & 92.99\% & 90.97\% \\
        Random          & 77.79\% & 46.33\% & 88.53\% & 60.83\% \\
        SC-Sim          & 93.42\% & 78.57\% & 91.08\% & 84.36\% \\
        WBFP-Eq         & 96.98\% & 86.39\% & 99.32\% & 92.41\% \\
         \textbf{WBFP-Wt}         & \textbf{97.39\%} & \textbf{88.64\%} & \textbf{99.36\%} & \textbf{93.69\%} \\
        \bottomrule
    \end{tabular}
\end{table}

\begin{table}[h!]
    \centering
    \caption{Comparison of Demonstration Selection Strategies for In-Context Learning with GPT-4 (Under Best Hybrid Strategy $\tau=300$).}
    \label{apptab:icl_results_large}
    \begin{tabular}{@{} l c c c c @{}}
        \toprule
        \textbf{Method} & \textbf{Accuracy} & \textbf{Precision} & \textbf{Recall} & \textbf{F1 Score} \\
        \midrule
        No-ICL          & 99.38\% & 100.00\% & 96.82\% & 98.38\% \\
        Random          & 89.00\% & 65.58\% & 90.97\% & 76.22\% \\
        SC-Sim          & 98.63\% & 100.00\% & 92.90\% & 96.32\% \\
        WBFP-Eq         & 99.62\% & 100.00\% & 98.03\% & 99.01\% \\
         WBFP-Wt        & \textbf{99.75\%} & \textbf{100.00\%} & \textbf{98.71\% }&\textbf{ 99.35\%} \\
        \bottomrule
    \end{tabular}
\end{table}

\end{document}